\documentclass[aps,pra,twocolumn,groupedaddress]{revtex4}

\usepackage{graphicx,color,rotating,amsmath,times}

\newcommand{\eqmulti}[1]{\begin{equation}\begin{split}#1\end{split}\end{equation}}

\newcommand{\ket}[1]{\ensuremath{\,|{#1}\rangle}}

\newcommand{\ketbra}[2]{\ensuremath{\,|{#1}\rangle\!\langle{#2}|\,}}

\newcommand{\op}[1]{\ensuremath{\mathrm{#1}}}

\newcommand{\imag}{{\bf i}}

\newcommand{\partdd}[1]{ \frac{\partial^2}{\partial #1^2} }

\newcommand{\trace}[2]{\textnormal{Tr}_{ #1 } \, #2\,}

\begin{document}

\title{Phase Diagram of Bosons in Two-Color Superlattices from Experimental Parameters}

\author{Felix Schmitt}
\author{Markus Hild}
\author{Robert Roth}

\affiliation{Institut f\"ur Kernphysik, Technische Universit\"at Darmstadt, 64289 Darmstadt, Germany}

\date{\today}

\begin{abstract}
We study the zero-temperature phase diagram of a gas of bosonic $^{87}$Rb atoms in two-color superlattice potentials starting directly from the experimental parameters, such as wavelengths and intensities of the two lasers generating the superlattice. In a first step, we map the experimental setup to a Bose-Hubbard Hamiltonian with site-dependent parameters through explicit band-structure calculations. In the second step, we solve the many-body problem using the density-matrix renormalization group (DMRG) approach  and compute observables such as energy gap, condensate fraction, maximum number fluctuations and visibility of interference fringes. We study the phase diagram as function of the laser intensities $s_2$ and $s_1$ as control parameters and show that all relevant quantum phases, i.e. superfluid, Mott-insulator, and quasi Bose-glass phase, and the transitions between them can be investigated through a variation of these intensities alone. 
\end{abstract}



\maketitle

\begin{figure}[b]
\includegraphics[width=\columnwidth]{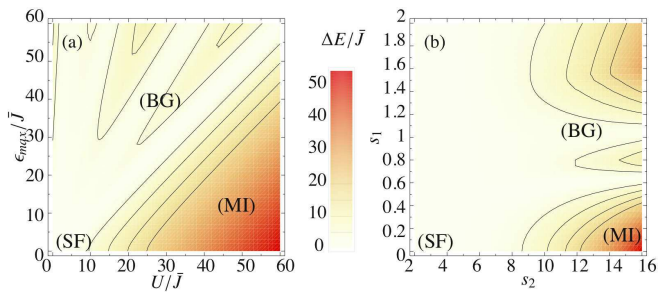}
\caption{(color online) Contour plots of the energy gap computed in DMRG for a commensurate superlattice with $I=N=30$ as function of the generic Hubbard parameters (a) and a function of the experimental laser intensities (b) (see text). The labels mark the domains of the superfluid (SF) phase, the homogeneous Mott-insulator (MI) phase, and the quasi Bose-glass (BG) phase.
\label{ComparePic}
}
\end{figure}

Ultracold atomic gases in optical lattice potentials offer unique possibilities for studying fundamental quantum phenomena, such as phase transitions to exotic quantum phases \cite{Greiner,Lye,Bloch,Fallani,Stoeferle,Billy,Roati}. Experimentally, the key advantages of these systems are the unparalleled in-situ control over all relevant parameters and the flexible tools to probe the quantum state. This has lead to a series of experimental studies, which recently also focus on non-homogeneous lattices and two-color superlattices \cite{Lye,Fallani,Billy,Roati}. Theoretically, these systems are well described by Hubbard-type Hamiltonians and a whole range of approaches from mean-field many-body methods to exact diagonalization and density-matrix renormalization group techniques is being used to study the phase diagram \cite{Fisher,Scalettar,RRoth0,RRoth1,Damski,Rapsch,Roux,Hild}. Theoretical studies typically adopt the generic parameters of the Bose-Hubbard Hamiltonian as control parameters, which entails several simplifications as compared to the experimental situation. A zero-temperature phase diagram for a Bose gas in a two-color superlattice in terms of the generic Hubbard parameters in shown in Fig. \ref{ComparePic}(a).

In this work we want to establish the direct link between the experimental setup as, e.g., discussed in Refs. \cite{Lye,Fallani} and the phase diagram by using the simplest experimental variables, the intensities $s_2$ and $s_1$ of the two lasers generating the superlattice, as control parameters. To this end, we first determine site-dependent Hubbard parameters from the experimental superlattice setup using explicit band-structure calculations. We then solve the many-body problem using state-of-the-art density-matrix renormalization group techniques and map out the phase diagram as function of the laser intensities. An example for such a phase diagram is shown in Fig. \ref{ComparePic}(b) and will be discussed in detail in the following.

\paragraph*{Bose-Hubbard Hamiltonian \& Band Structure.} 

Strongly correlated ultracold bosonic atoms in a sufficiently deep one-dimensional optical lattice potential with $I$ sites are well described by the single-band Bose-Hubbard Hamiltonian \cite{Jaksch}
\eqmulti{ \label{HubbardHamil}
	\op{H} 
	= \sum_{i=1}^{I} \Big\{ -J_{i,i+1}  ( \op{a}_{i+1}^{\dagger} \op{a}_i^{} + \text{h.a.}) 
    + \frac{U_i}{2} \op{n}^{}_i (\op{n}^{}_i-1 )     
    +\epsilon_i \op{n}^{}_i \Big\} \,.
}
Here $\op{a}^{\dagger}_i$ ($\op{a}^{}_i$) are creation (annihilation) operators for a boson in the lowest Wannier state localized at site $i$ and  $\op{n}_i^{}=\op{a}^{\dagger}_i \op{a}^{}_i$ are the corresponding occupation number operators. The three terms of the Hamiltonian describe the tunneling between adjacent sites, the on-site two-body interaction and the on-site potential, respectively. The site-dependent Hubbard parameters $J_{i,i+1}$, $U_i$, and $\epsilon_i$, which control the strength of the individual terms, contain all information on the underlying lattice potential and the interaction between the atoms. 

In most previous applications, the Hubbard parameters were used directly as the control parameters spanning the phase diagram. Superlattice systems are typical approximated by using site-independent tunneling and interaction matrix elements, i.e. $J_{i,i+1}\equiv J$ and $U_i\equiv U$, and some ansatz for the site-dependent on-site energies $\epsilon_i$ \cite{RRoth0,RRoth1}. 
In this study we start directly from the experimental setup, i.e., from a one-dimensional lattice potential $V(x)$ in real-space depending on the wavelengths $\lambda_1$ and $\lambda_2$ and the intensities $s_1$ and $s_2$ of the two standing waves through 
$V(x) 
=  s_1  E_{r_1} \sin^2\big(\frac{2 \pi}{\lambda_1} x + \phi \big) 
+ s_2  E_{r_2} \sin^2\big(\frac{2 \pi}{\lambda_2}x\big)$,
where $E_{r_i}=\frac{h^2}{2 m \lambda_i^2}$ is the recoil energy of the atoms with mass $m$ and $\phi$ a relative phase shift between the two standing waves. For the sake of simplicity we do not account for a longitudinal trapping potential. We study two different superlattice topologies: A commensurate superlattice with $\lambda_2=800$ nm, $\lambda_1=1000$ nm ($\phi=\pi/4$) and an incommensurate superlattice with $\lambda_2=830$ nm, $\lambda_1=1076$ nm ($\phi=\pi/3$). These values are motivated by the experimental setup discussed in Ref. \cite{Lye} and we also assume the strong primary laser to be defined by $\lambda_2$ and $s_2$.

In order to extract the Hubbard parameters for the lattice potential $V(x)$ we first have to determine the localized Wannier functions via a single-particle band-structure calculation. For a homogeneous lattice ($s_1=0$) with $I$ sites we can easily obtain a numerical solution for the Bloch functions of the lowest band, $\psi_{k}(x)$, characterized by a quasimomentum index $k=0, 1,\dots, I-1$. The corresponding Wannier functions are then determined by a Fourier transformation with respect to the quasimomentum, $w_{i}(x)= \frac{1}{\sqrt{I}} \sum_{k=0}^{I-1} e^{-\imag \frac{2\pi}{I} k i}\, e^{\imag \varphi_k}\, \psi_{k}(x)$. The arbitrary phase $\varphi_k$ of the individual Bloch functions is determined by requiring the Wannier functions to be  localized. From these Wannier functions the Hubbard parameters can be computed through matrix elements of the different terms of the real-space Hamiltonian
\eqmulti{ \label{eq:hubbardparam}
  -J_{i,i+1} 
  &= \int dx \;  w_{i}^{\ast}(x) \left( -\frac{\hbar^2}{2m} \partdd{x} + V_{}(x) \right)  w_{{i+1}}(x) \\
  \epsilon_{i}  
  &= \int dx \;  w_{i}^{\ast}(x) \left( -\frac{\hbar^2}{2m} \partdd{x} + V_{}(x) \right)  w_{i}(x) \\
  U_i  
  &=  2 \hbar\omega_{\perp}\, a_s\,  \int dx \; |w_{i}(x)|^4 \,.
}
The interaction is described by a three-dimensional s-wave contact interaction with a strength given by the s-wave scattering length $a_s$. For the transverse directions the wave function is approximated by a Gaussian of a width corresponding to the frequency $\omega_{\perp}$ of the transverse confinement. Following Ref. \cite{Lye,Fallani} we consider a gas of $^{87}$Rb atoms with s-wave scattering length $a_s = 109 a_{\text{Bohr}}$ and we assume a transverse trapping frequency $\omega_{\perp}=30 E_{r_2}/h$ to ensure the validity of the 1D description.

In the case of non-homogeneous lattice potentials we describe each site individually with a Wannier function extracted from a homogeneous lattice with the same local depths. Based on this set of nontrivial site-dependent Wannier functions the Hubbard parameters are then determined from Eqs. \eqref{eq:hubbardparam}. This scheme is sufficiently accurate for the parameter regime under consideration. Even simpler approximations, e.g. a lowest-order perturbative inclusion of the lattice inhomogeneity by computing the Hubbard parameters \eqref{eq:hubbardparam} with the full potential $V(x)$ and the Wannier functions from a homogeneous lattice of the same average depth, yield similar results. 

\begin{figure}
\includegraphics[width=0.75\columnwidth]{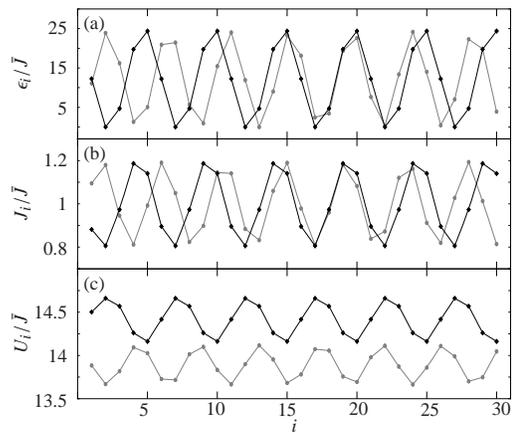}
\caption{Site-dependent Hubbard parameters obtained from band-structure calculations for two-color superlattices with commensurate (black symbols) and incommensurate wavelengths (gray symbols) for $s_2=10$ and $s_1=1$. Lines to guide the eye.}
\label{ParameterPic}
\end{figure}

Examples for the site-dependent Hubbard parameters for the commensurate as well as the incommensurate superlattice with $s_2=10$ and $s_1=1$ are depicted in Fig. \ref{ParameterPic}. Note that we always subtract a global constant from the Hamiltonian such that $\epsilon_{\text{min}}=0$. The dominant effect of the superlattice structure is the spatial modulation of the on-site energies $\epsilon_{i}$, which is in-line with the approximation to introduce the superlattice through $\epsilon_i$ only. However, also the tunneling matrix element $J_{i,i+1}$, which essentially depends on the barrier height between the sites $i$ and $i+1$, shows a sizable variation of $\pm20\%$ around the average value $\bar{J}$. The interaction strength $U_i$ shows only a weak site-dependence which is induced solely through the site-dependence of the Wannier functions. A comparison of the energy scales reveals that the weak secondary laser with $s_1 = s_2/10$ considered in this example is sufficient to create a superlattice with $\epsilon_{\text{max}} >\bar{U}$. For the following calculations we consider a parameter range $s_2=2$ to $16$ and $s_1=0$ to $2$. For fixed $s_1=0$, i.e. in a homogeneous lattice, the variation of $s_2$ covers a range from $U/J\approx 1$ to $60$ with $\epsilon_{\text{max}}/\bar{J}=0$. The variation of $s_1$ spans a range up to $\epsilon_{\text{max}}/\bar{J}\approx4$ at $s_2=2$ and up to $\epsilon_{\text{max}}/\bar{J}\approx180$ at $s_2=16$ with $\bar{U}/\bar{J}$ being almost independent of $s_1$. Throughout this work the bars indicate averages over all lattice sites.
 
\paragraph*{Density Matrix Renormalization Group \& Observables.} 

Based on the Bose-Hubbard Hamiltonian with the site-dependent parameters discussed above we solve the many-particle problem using the density-matrix renormalization group (DMRG) approach \cite{White,Schollwoeck}, which is one of the most powerful methods for solving one-dimensional problems of this type. 

The so-called infinite-size DMRG algorithm is based on an iterative growth of the lattice system. We start from an initial block of $I_b$ lattice sites with up to $N_{\text{b}}$ particles per site described in a Fock space $\mathcal{F}_{\text{b}}$ of dimension $D_{\text{b}}$. To this block an additional site with up to $N_{\text{s}}$ particles described by a space $\mathcal{F}_{\text{s}}$ is attached to build the system space $\mathcal{F}_{\text{sys}}=\mathcal{F}_{\text{b}}\otimes\mathcal{F}_{\text{s}}$ of dimension $D_{\text{sys}}=D_{\text{b}} D_{\text{s}}$. In order to mimic the thermodynamic limes, the system is coupled with an analogously constructed environment yielding the superblock $\mathcal{F}_{\text{super}}=\mathcal{F}_{\text{sys}}\otimes\mathcal{F}_{\text{env}}$ with the constraint of fixed total particle number. The ground state $\ket{\psi_0}$ is obtained by diagonalizing the superblock Hamiltonian. A reduced density-matrix is formed by tracing out the environment, $\rho_{\text{red}}=\trace{\text{env}}{\ketbra{\psi_0}{\psi_0}}$, and the $D_{\text{b}}$ eigenvectors of $\rho_{\text{red}}$ for the largest eigenvalues are used to span the Fock space $\tilde{\mathcal{F}}_{\text{b}}$ of a new block of length $\tilde{I}_{\text{b}}=I_{\text{b}}+1$ for the next iteration. These eigenvectors also define a non-unitary transformation matrix $\mathcal{O}$ which is employed to construct the new block Hamiltonian $\tilde{H}_{\text{b}}=\mathcal{O}^{\dagger} H_{\text{sys}} \mathcal{O}$. This cycle is repeated until the final length of the lattice is reached. 

In non-homogeneous systems only at the very last step of the algorithm the full information about the lattice topology is included. Therefore, the infinite-size algorithm alone is not sufficient to provide an accurate description of the ground state for superlattices. During a second stage of the calculation we thus resort to the so-called finite-size DMRG algorithm. 
Keeping the size of the superblock fixed at its final value, the size of the system is increased at the expense of the environment and vice versa. While sweeping the boundary between system and environment forth and back using the same operations discussed for the infinite-size algorithm, the Hamiltonian always takes the whole lattice into account. 

For the following calculations we use the inifinite-size algorithm for the initial phase and perform three complete sweeps of the finite-size algorithms in the second phase. To warrant convergence of the whole procedure, we increase of the block basis dimension $D_{\text{b}}$ until all observables stabilize. We typically use $D_{\text{b}}=56$ with up to five particles per lattice site $N_{\text{b}}=N_{\text{s}}=5$. This restriction is sufficient because we are most interested in the strongly correlated regime. Among the observables we analyze are the energy gap $\Delta E$ between ground and first excited state, the condensate fraction $f_{\text{c}}$ given by the largest eigenvalue of the one-body density matrix, the maximum fluctuation $\sigma_{\text{max}}$ of the occupation numbers across the lattice, and the visibility $\nu$ of the interference fringes extracted from the quasimomentum occupation numbers. The latter is of particular interest as it is directly accessible in experiment. The detailed definition of those observables is discussed in Refs. \cite{RRoth1,RRoth2,RRoth3}. 

\begin{figure*}
\includegraphics[height=0.35\textheight]{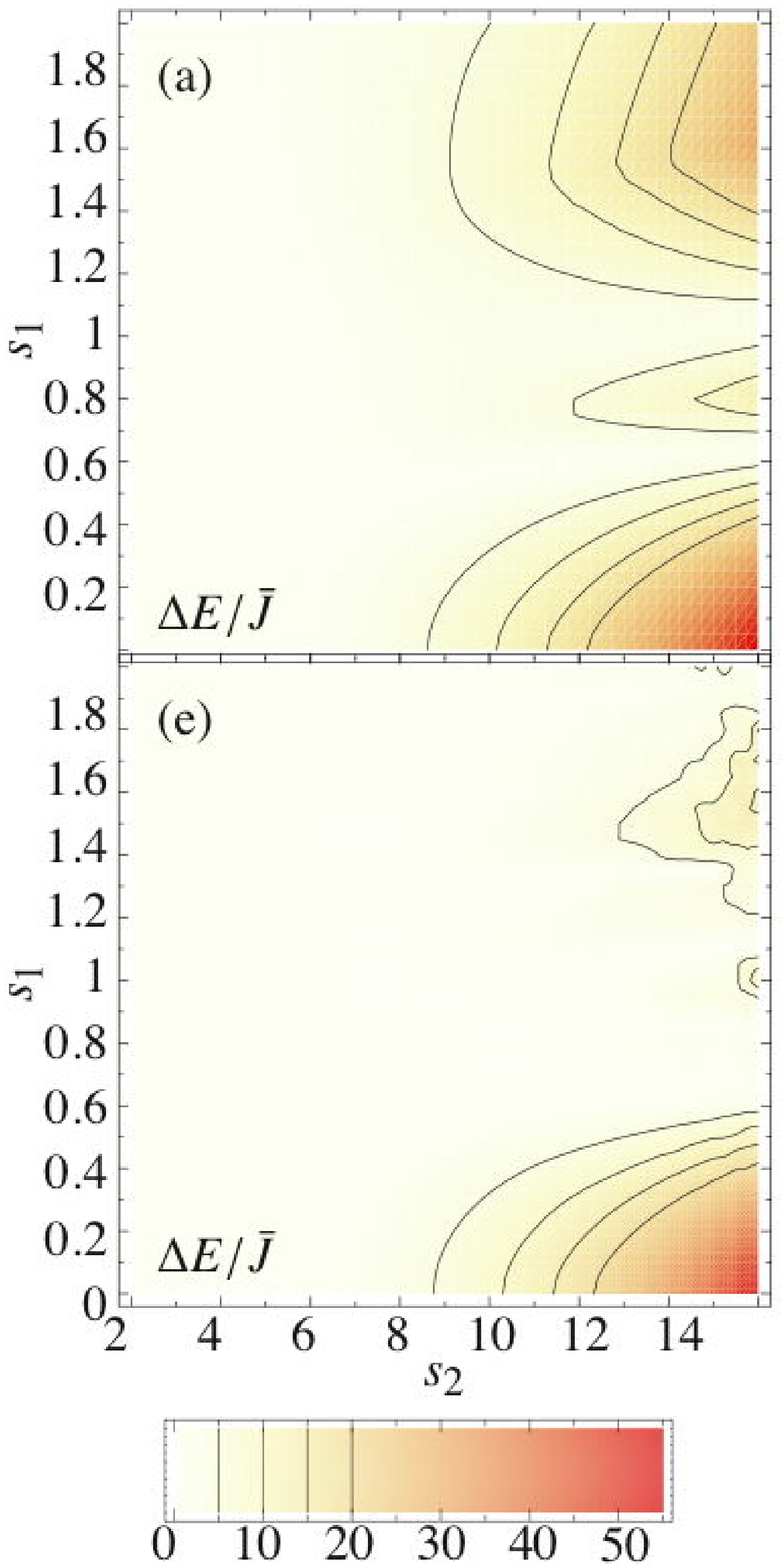}%
\includegraphics[height=0.35\textheight]{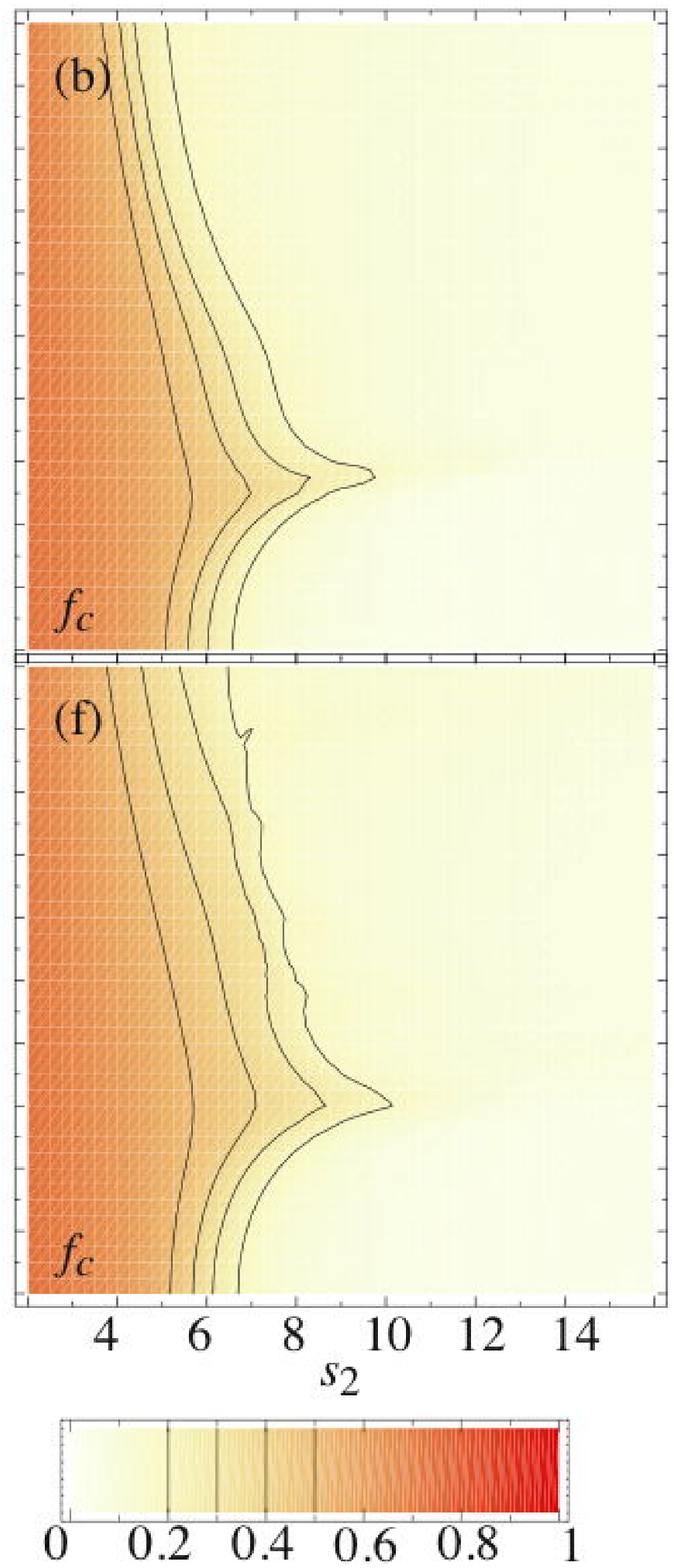}%
\includegraphics[height=0.35\textheight]{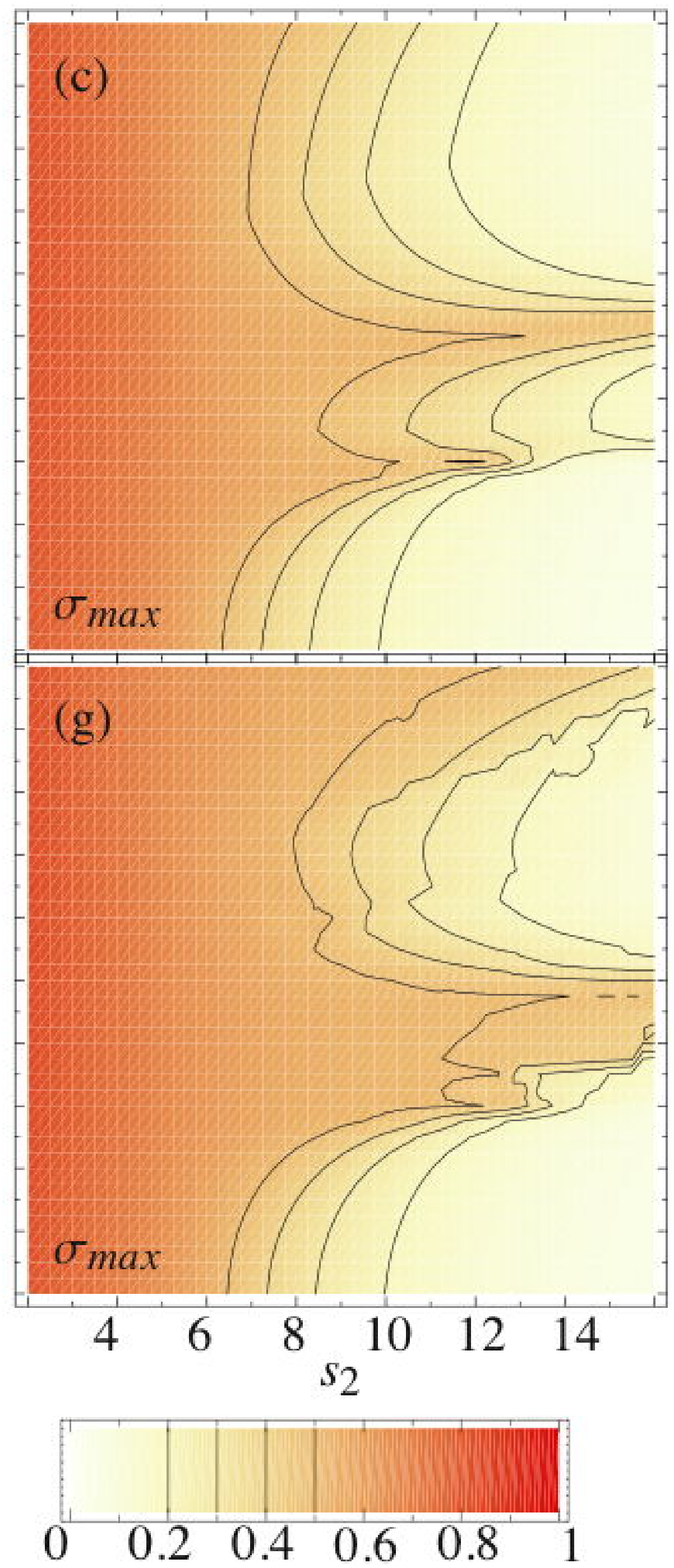}%
\includegraphics[height=0.35\textheight]{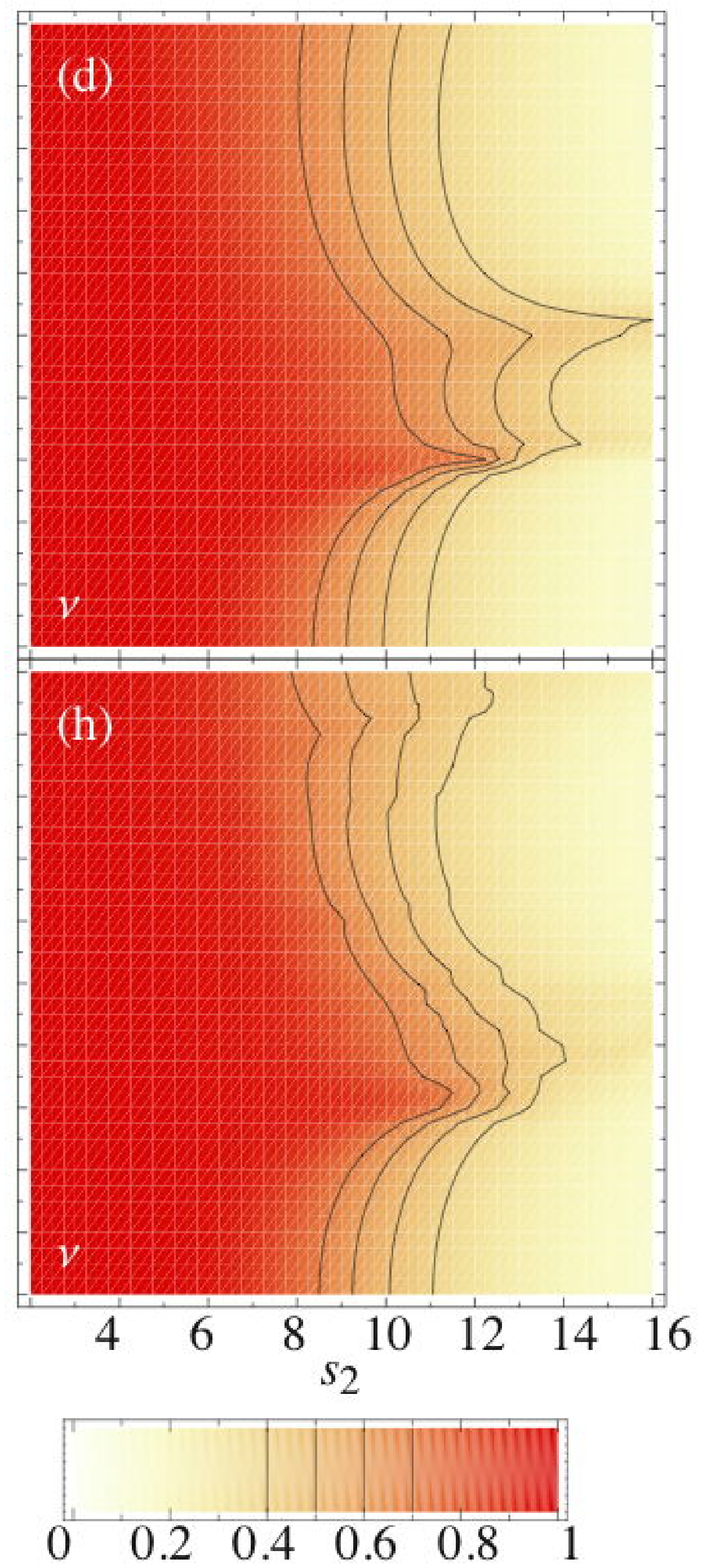}%
\caption{(color online) Phase diagrams in terms of energy gap $\Delta E/\bar{J}$, condensate fraction $f_{\text{c}}$, maximum number fluctuations $\sigma_{\text{max}}$, and visibility $\nu$ obtained from DMRG calculations for $N=I=30$ with the commensurate superlattice (upper row) and the incommensurate superlattice (lower row). The irregularities of the contour lines are due to minimal numerical fluctuations of the DMRG results.}
\label{Pic}
\end{figure*}

\paragraph*{Commensurate Lattice.}

We first consider the commensurate superlattice ($\lambda_2=800$ nm, $\lambda_1=1000$ nm, $\phi=\pi/4$) with the site-dependent Hubbard-parameters depicted in Fig. \ref{ParameterPic}. Already in Fig. \ref{ComparePic} we have used this superlattice to illustrate the difference between a generic phase diagram spanned by the the Hubbard parameters $U/J$ and $\epsilon_{\text{max}}/J$ (neglecting the site-dependence of $U$ and $J$) and the experiment-specific phase diagram spanned by the laser intensities $s_2$ and $s_1$. In both cases we show the energy gap obtained from DMRG calculations for $I=N=30$. Because the variation of $s_2$ and $s_1$ affects all site-dependent Hubbard parameters simultaneously, the $(s_2,s_1)$ phase diagram is distorted relative to the generic $(U/J,\epsilon_{\text{max}}/J)$ phase diagram. Nonetheless, all relevant quantum phases can be reached by variation of the laser intensities alone, i.e., there is no need to vary other experimental parameters like the interatomic scattering length.

A more detailed analysis of the phase diagram as function of $s_2$ and $s_1$ is given in Figs. \ref{Pic}(a) to (d), where we depict the energy gap, the condensate fraction, the maximum number fluctuations, and the visibility, resp., obtained in DMRG calculation for $I=N=30$ with the commensurate superlattice. 
The superfluid (SF) phase is signaled by a small or vanishing energy gap, a large condensate fraction, large number fluctuations, and maximum visibility. Although the most stringent order parameter for the SF phase---the superfluid fraction \cite{RRoth2}---is not computed here, this signature allows us to identify the SF phase in the regime of small $s_2$, roughly up to $s_2\lesssim6$ for all $s_1\lesssim2$. In this regime the superlattice is shallow such that tunneling dominates over on-site interactions, which is a prerequisite for the long-range coherence present in the SF phase. For $s_1=0$ the ratio $U/J$ is about $1$ at $s_2=2$ and reaches about $4.5$ at $s_2=6$. Around this value of $U/J$ the phase transition from superfluid to Mott insulator is expected in one-dimensional Bose systems in homogeneous lattices \cite{Roux,RRoth2}, which is consistent with our observation. Even in the presence of the secondary laser, i.e. for $0<s_1\leq2$, the intensity $s_2=6$ leads to $\bar{U}/\bar{J}\approx4.5$, which explains the presence of  the SF phase in the whole range of $s_1$ considered here. 

If we increase $s_2$ at fixed $s_1=0$, the system enters the homogeneous Mott-insulator (MI) phase, which is characterized by a large energy gap, a small condensate fraction, small fluctuations, and a minimal visibility. These signatures are clearly visible in the Fig. \ref{Pic} for large values of $s_2$ and small values of $s_1$. At $s_1=0$ and $s_2=16$ the ratio $U/J$ is about $60$ and the system is deep in the MI regime.

If we now increase $s_1$ at fixed $s_2=16$, the modulation of the site-dependent Hubbard parameters grows rapidly. Already at $s_1\approx0.6$ the spread of the on-site energies becomes comparable to the average interaction strength, i.e., $\epsilon_{\text{max}}/\bar{J}\approx\bar{U}/\bar{J}\approx60$. Thus it becomes energetically favorable to move particles from the lattice sites with largest on-site energies to the sites with the lowest on-site energies, thus creating doubly occupied sites. In this way the homogeneous MI phase is broken up and the transition to a quasi Bose glass (BG) phase is observed. In a truely random infinite-size lattice, the BG phase is characterized by a vanishing energy gap. Intuitively, this results from the continuous distribution of on-site energies, which allows for the construction of excited states by redistributions of atoms to sites with infinitesimally higher on-site energies and thus infinitesimal excitation energies. In contrast to a random lattice, the commensurate superlattice exhibits only 5 different on-site energies. This always gives rise to finite energy gaps and extended domains in the phase diagram, where certain occupation-number states dominate the ground state---two of those domains are visible in Fig. \ref{Pic}(a). Only in the transition regions between those domains the energy gaps become small. In order to approach a more realistic BG phase, we have to consider more complex lattice topologies, e.g., the incommensurate superlattice.

\paragraph*{Incommensurate Lattice.} 

We repeat the above analysis for the incommensurate lattice ($\lambda_2=830$ nm, $\lambda_1=1076$ nm, $\phi=\pi/3$) inspired by the experimental setup of Lye et al. \cite{Lye,Fallani}. These parameters lead to a modulation of the site-dependent Hubbard parameters with a periodicity which does not correspond to an integer number of lattice sites, as seen in Fig. \ref{ParameterPic}. Therefore, the pattern of on-site energies $\epsilon_{i}$, for example, is not periodic anymore as in the case of the commensurate lattice.

The resulting phase diagrams for $I=N=30$ are depicted in Figs. \ref{Pic}(e) to (h). Evidently, the structure and extend of the SF and the MI phase are not affected by the change in the lattice topology. Only the BG phase exhibits a different behavior. The energy gap is clearly reduced due to the irregular character of the superlattice. The larger number of different on-site energies allows for redistributions associated with lower excitation energies and reduced energy gaps. The more irregular the lattice the stronger the reduction of the energy gap in the quasi BG phase---eventually the true BG phase in a random lattice would be approached. Observables like the condensate fraction and the visibility are still suppressed in the BG regime and allow for a unique distinction from the SF phase.

\paragraph*{Conclusions.} 

We have studied the phase diagram of bosonic atoms in one-dimensional two-color superlattices starting from the experimentally accessible parameters of the optical lattice. Following a band-structure calculation to extract the corresponding site-dependent Hubbard parameters the many-body problem is solved using the DMRG approach. Using the intensities of the two laser fields, $s_1$ and $s_2$, as control parameters while keeping all other experimental parameters fixed, it is possible to access all relevant quantum phases of the system. By following a simple path in the $(s_1,s_2)$-parameter-plane, the transition from SF to MI phase (variation of $s_2$ for fixed $s_1\lesssim 0.4$) and the transition from MI to quasi BG phase (variation of $s_1$ for fixed $s_2\gtrsim10$) can be studied in detail. By comparing superlattices obtained from lasers with commensurate and incommensurate wavelengths we show that the SF and the MI phase are largely insensitive to the detailed lattice topology. As expected, only the energy gaps in the quasi BG phase change systematically when going to a more irregular lattice topology and approach the limit of a gapless BG expected in a random lattice. By measuring the energy gap and, e.g., the visibility of the interference fringes it is possible to uniquely identify the three different quantum phases in experiment. Thus a detailed experimental analysis of the rich structure of the phase diagram is directly possible, both from the control and the diagnostics point of view.

Supported by the Helmholtz Alliance HA216-TUD/EMMI.


\end{document}